\documentclass[12pt]{iopart}
\usepackage{iopams}
\usepackage{graphicx}
\usepackage{latexsym}
\usepackage{epsf}
\usepackage{epsfig}
\begin{document}

\title{Time-Varying Lense-Thirring System}

\author{C. Chicone$^1$ and B. Mashhoon$^2$}

\address{$^1$ Department of Mathematics, University of Missouri-Columbia, Columbia, MO 65211, USA}

\address{$^2$ Department of Physics and Astronomy, University of Missouri-Columbia, Columbia, MO 65211, USA}

\begin{abstract}
We consider bound geodesic orbits of test masses in the exterior gravitational field of a rotating astronomical source whose proper angular momentum varies linearly with time. The linear perturbation approach of Lense and Thirring is herein extended to the nonstationary case. In particular, we investigate the instability of Lense-Thirring precessing orbits due to the slow temporal variation of the gravitomagnetic field of the source.
\end{abstract}


\section{Introduction}\label{s1}

Nine decades ago, Lense and Thirring considered the motion of a free test mass in the stationary exterior gravitational field of a rotating astronomical source within the framework of general relativity \cite{1,2}. They treated the influence of the gravitomagnetic field on the particle orbit via the Lagrange planetary equations and showed by means of linear perturbation theory that---when averaged over the fast Keplerian motion---the orbit keeps its shape and slowly precesses. Specifically, they found that the semimajor axis $a$ of the osculating ellipse does not change, while its eccentricity $e$ and orbital inclination $i$ contain periodic terms that vanish on the average; moreover, the osculating ellipse precesses.  This occurs both within the orbital plane of the osculating ellipse and without, as the orbital plane precesses about the axis of rotation of the central source. The frequency of both precessions can be described by
\begin{equation}\label{eq:1}
\boldsymbol{\omega}_{LT}=\frac{2G}{a^3(1-e^2)^{3/2}} [\mathbf{J}_0-3 (\mathbf{J}_0\cdot \hat{\mathbf{n}}) \hat{\mathbf{n}} ].\end{equation}
Here, $\mathbf{J}_0$ is the constant angular momentum of the source, $\hat{\mathbf{n}}$ is a unit vector parallel to the orbital angular momentum of the osculating ellipse and $\hat{\mathbf{J}}_0 \cdot \hat{\mathbf{n}}=\cos i$. Thus the Runge-Lenz vector and the orbital angular momentum vector of the osculating ellipse both precess with the Lense-Thirring frequency \eref{eq:1}.

Astronomical bodies in general rotate; however, the magnitude of the proper angular momentum is seldom constant. In two recent papers \cite{3,4}, the gravitational physics around a rotating central source whose spin angular momentum vector is fixed in space but varies linearly in time has been explored. In particular, it has been shown in \cite{3} that sufficiently far from such a source, the spacetime metric is given by
\begin{equation}\label{eq:2}
ds^2=-c^2 \left( 1-2\frac{\Phi}{c^2}\right) dt^2-\frac{4}{c}(\mathbf{A}\cdot d\mathbf{x}) dt+\left( 1+2\frac{\Phi}{c^2} \right) \delta_{ij}dx^idx^j,
\end{equation}
where
\begin{equation}\label{eq:3}
\Phi =\frac{GM}{r},\quad \mathbf{A}=\frac{G}{c} \frac{\mathbf{J}(t)\times \mathbf{x}}{r^3}
\end{equation}
are the gravitoelectric and gravitomagnetic potentials, respectively. Here, $r=|\mathbf{x}|$, $M$ is the mass of the source and its angular momentum is given by
\begin{equation}\label{eq:4}
\mathbf{J}(t)=(J_0+J_1t)\hat{\mathbf{z}};
\end{equation}
moreover, $\Phi\ll c^2$ and $|\mathbf{A}|\ll c^2$. Thus $r\gg GM/c^2$, $r\gg J/(Mc)$ and all $O(c^{-4})$ contributions to the metric tensor have been neglected in this linear post-Newtonian approach to general relativity. As explained in \cite{3,4}, we simply ignore the processes by which the
variation of angular momentum is turned on and off and assume that equation
\eref{eq:4} holds throughout the temporal interval of interest; furthermore, all
radiative effects are neglected.

The motion of a free test particle in the gravitational field of the source is given by the geodesic equation in the spacetime with metric \eref{eq:2}. This equation, as shown in \cite{4}, can be written in its reduced form
\begin{eqnarray}\label{eq:5}
 \nonumber \fl \frac{d\mathbf{v}}{dt} +\frac{GM\mathbf{x}}{r^3} =& \frac{GM}{c^2r^3} [4(\mathbf{x}\cdot \mathbf{v} )\mathbf{v}-v^2\mathbf{x}] + \frac{2G}{c^2} \frac{\dot{\mathbf{J}} \times \mathbf{x}}{r^3} -\frac{2}{c}\mathbf{v} \times \mathbf{B}\\ &{}-\frac{6GJ(t)}{c^4r^5} [\hat{\mathbf{J}} \cdot (\mathbf{x}\times \mathbf{v})] (\mathbf{x}\cdot \mathbf{v})\mathbf{v},
\end{eqnarray}
where an overdot represents differentiation with respect to time $t$ and $\mathbf{B}=\boldsymbol{\nabla}\times \mathbf{A}$ is the gravitomagnetic field given by
\begin{equation}\label{eq:6}
\mathbf{B}=\frac{G(J_0+J_1t)}{cr^5} (3z\mathbf{x} -r^2\hat{\mathbf{z}}).
\end{equation}

The right-hand side of \eref{eq:5} contains all of the linear post-Newtonian contributions that arise from potentials given in \eref{eq:3}. It turns out, however, that in a general treatment to $O(c^{-2})$, the nonlinear gravitoelectric term $4G^2M^2\mathbf{x}/(c^2r^4)$, which is quadratic in $\Phi$ and hence absent in our linear treatment, should also be taken into account. In the present work, we explore further the influence of the temporal variation of $J$ on motion around a central rotating source to first post-Newtonian order, namely, $O(c^{-2})$. Thus instead of \eref{eq:5}, we consider
\begin{eqnarray}\label{eq:7} 
 \frac{d\mathbf{v}}{dt}+\frac{GM\mathbf{x}}{r^3} =\mathbf{F},\\
\mathbf{F} =\frac{GM}{c^2r^3} [4(\mathbf{x}\cdot \mathbf{v})\mathbf{v}-v^2\mathbf{x}] + \frac{4G^2M^2}{c^2r^4} \mathbf{x}+\frac{2G}{c^2} \frac{\dot{\mathbf{J}}\times \mathbf{x}}{r^3} -\frac{2}{c}\mathbf{v}\times \mathbf{B}.
\label{eq:8}\end{eqnarray}

As demonstrated in \cite{3,4}, equation \eref{eq:2} represents the metric of a nonstationary linearized Kerr spacetime. The geodesic equation in Kerr spacetime is completely integrable \cite{5}; more recent results are contained, for instance, in \cite{6} and references therein.

\section{Instability of spherical orbits}\label{s2}

To gain insight into the nature of allowed orbits, it proves useful to study first certain simple configurations. For instance, it has been shown in \cite{4} that \emph{circular} equatorial orbits are unstable due to the temporal variation of $J$. We therefore look for \emph{spherical} orbits in this section. It will turn out that the post-Newtonian equations of motion \eref{eq:7} and \eref{eq:8} do not allow spherical orbits unless $\dot{J}=0$. For $\dot{J}=0$, the spherical orbits can be concisely characterized as describing a post-Keplerian circular orbit undergoing Lense-Thirring precession. We will then investigate the instability of spherical orbits under the slow temporal variation of $J$.

\subsection{Equations of motion in spherical coordinates}\label{ss2.1}

To investigate spherical orbits, it is convenient to write \eref{eq:7} and \eref{eq:8} in spherical polar coordinates $(r,\theta ,\phi)$ such that $x=r\sin \theta \cos \phi$, $y=r\sin \theta \sin \phi$ and $z=r\cos \theta$. In terms of the corresponding unit vectors $\hat{\mathbf{r}}$, $\hat{\boldsymbol{\theta}}$ and $\hat{\boldsymbol{\phi}}$, one can write $\mathbf{x}=r\hat{\mathbf{r}}$,
\begin{eqnarray}\label{eq:9}
 \dot{\mathbf{x}}=\mathbf{v}=\dot{r} \hat{\mathbf{r}} +r\dot{\theta} \hat{\boldsymbol{\theta}} +r \dot{\phi}\sin \theta\, \hat{\boldsymbol{\phi}},\\
\label{eq:10} \eqalign{\ddot{\mathbf{x}} =& (\ddot{r} -r\dot{\theta}^2-r \dot{\phi}^2 \sin^2\theta )\hat{\mathbf{r}} +(r\ddot{\theta} +2\dot{r}\dot{\theta} -r\dot{\phi}^2 \sin \theta \cos \theta ) \hat{\boldsymbol{\theta}}\cr
&{} +(r\ddot{\phi}\sin \theta  +2\dot{r}\dot{\phi}\sin \theta  +2r\dot{\theta} \dot{\phi} \cos \theta )\hat{\boldsymbol{\phi}}.
\label{eq:10.1}}\end{eqnarray}
Moreover, it follows from \eref{eq:8} that $\mathbf{F}=F_r\hat{\mathbf{r}} +F_\theta \hat{\boldsymbol{\theta}} +F_\phi \hat{\boldsymbol{\phi}}$, where
\begin{eqnarray}\label{eq:11}
F_r =\frac{GM}{c^2r^2} \left(4\dot{r}^2-v^2+\frac{4GM}{r}\right) +\frac{2GJ(t)}{c^2r^2}\dot{\phi}\sin ^2\theta ,\\
F_\theta =\frac{4GM}{c^2r} \dot{r}\dot{\theta} -\frac{4GJ(t)}{c^2r^2}\dot{\phi} \sin \theta \cos \theta ,\label{eq:12}\\
F_\phi = \frac{4GM}{c^2r}\dot{r}\dot{\phi} \sin \theta  +\frac{2G\dot{J}} {c^2r^2} \sin \theta -\frac{2GJ(t)}{c^2r^3} (\dot{r}\sin \theta -2r\dot{\theta}\cos \theta ).
\label{eq:13}\end{eqnarray}
Thus the post-Newtonian equations of motion are
\begin{eqnarray}\label{eq:14}
\fl \ddot{r}-r\dot{\theta}^2 -r\dot{\phi}^2\sin ^2 \theta +\frac{GM}{r^2} =\frac{GM}{c^2r^2} \left( 4\dot{r}^2-v^2+\frac{4GM}{r}\right) 
+\frac{2GJ(t)}{c^2r^2} \dot{\phi} \sin ^2\theta ,\\
\fl r\ddot{\theta} +2\dot{r}\dot{\theta} -r\dot{\phi}^2\sin \theta \cos \theta  =\frac{4GM}{c^2r} \dot{r}\dot{\theta} -\frac{4GJ(t)}{c^2r^2} \dot{\phi} \sin \theta \cos \theta ,\label{eq:15}\\
\nonumber \fl r\ddot{\phi}\sin \theta +2\dot{r}\dot{\phi} \sin \theta  +2r \dot{\theta} \dot{\phi} \cos \theta=\frac{4GM}{c^2r} \dot{r}\dot{\phi}\sin \theta   
+\frac{2G\dot{J}}{c^2r^2}\sin \theta\\ \hspace{1.5in}-\frac{2GJ(t)}{c^2r^3} (\dot{r}\sin \theta -2r\dot{\theta}\cos \theta ).\label{eq:16}
\end{eqnarray}

It is interesting to note that \eref{eq:16} may be written as
\begin{equation}\label{eq:17}
\frac{d}{dt} \left[ r^2\dot{\phi} \sin ^2\theta  -\frac{2GJ(t)}{c^2r} \sin ^2\theta \right] =\frac{4GM}{c^2} r\dot{r}\dot{\phi}\sin ^2\theta .
\end{equation}
Thus an integral of the motion is obtained if the right-hand side of \eref{eq:17} vanishes. We therefore look for spherical orbits $(r,\theta ,\phi)=(\rho ,\vartheta ,\varphi)$ such that $\rho$ is a constant.

\subsection{Spherical orbits}\label{ss2.2}

It follows from \eref{eq:17} that for a spherical orbit $(\rho ,\vartheta ,\varphi)$,
\begin{equation}\label{eq:18}
\dot{\varphi} =\frac{C}{\sin ^2\vartheta} +\frac{2GJ(t)}{c^2\rho^3},
\end{equation}
where $C$ is a constant of integration. Substituting \eref{eq:18} in \eref{eq:14} and \eref{eq:15}, and keeping terms only up to $O(c^{-2})$, we find
\begin{equation}\label{eq:19}
\dot{\vartheta} ^2+\frac{C^2}{\sin^2\vartheta}=\frac{GM}{\rho^3} \left(1-\frac{3GM}{c^2\rho}\right) - \frac{6GCJ(t)}{c^2\rho^3},
\end{equation}
and
\begin{equation}\label{eq:20}
\ddot{\vartheta} -\frac{C^2\cos \vartheta}{\sin ^3\vartheta} =0,
\end{equation}
respectively. These equations are compatible only if $\dot{J}=0$. Thus let $J=J_0$ and define the positive constant $\Omega >0$ such that
\begin{equation}\label{eq:21}
\Omega^2 =\frac{GM}{\rho^3} \left( 1-\frac{3GM}{c^2\rho} \right) -\frac{6GJ_0}{c^2\rho^3} C.
\end{equation}

To find the motion in $\vartheta$, we note that \eref{eq:19} can be written as
\begin{equation}\label{eq:22}
\left( \frac{d\cos \vartheta}{dt}\right)^2 =(\Omega^2-C^2)-\Omega^2\cos ^2\vartheta,
\end{equation}
which has solutions only for $\Omega^2\geq C^2$; in this case, we have
\begin{equation}\label{eq:23}
\cos \vartheta =\alpha \sin (\Omega t+\beta ),
\end{equation}
where $\alpha$ is given by
\begin{equation}\label{eq:24}
C^2=\Omega^2 (1-\alpha^2)
\end{equation}
and $\beta$ is a constant phase. It is possible to let $C=\Omega \cos i$ and $\alpha =\pm \sin i$, where $i$ is a constant (``inclination'') angle. Then the solution of \eref{eq:18} may be expressed as
\begin{equation}\label{eq:25}
\varphi (t)=\frac{2GJ_0t}{c^2\rho^3} +\tan^{-1} [\cos i\tan (\Omega t+\beta )]+\varphi_0,
\end{equation}
where $\varphi_0$ is an integration constant. Let us note that with 
\begin{equation}\label{eq:26}
\omega_K=\left( \frac{GM}{\rho^3}\right)^{1/2}, \quad \omega=\omega_K\left( 1-\frac{3}{2}\frac{GM}{c^2\rho}\right),
\end{equation}
where $\omega_K$ is the Keplerian frequency, we have
\begin{equation}\label{eq:27}
\Omega =\omega -\frac{3GJ_0\cos i}{c^2\rho^3} .
\end{equation}
Only positive square roots are considered throughout. The explicit solution for spherical orbits in the first post-Newtonian approximation is contained in \eref{eq:23} and \eref{eq:25}. Spherical orbits in Kerr spacetime have been considered in \cite{7}.

It is interesting to note that when $J_0=0$ in \eref{eq:23} and \eref{eq:25}, a spherical orbit simply reduces to a circular orbit of radius $\rho$ and frequency $\omega$ about mass $M$ such that the orbital angular momentum vector makes an angle of $i$ with the $z$ axis. Thus one can characterize the spherical orbits under consideration as circular orbits in the post-Newtonian gravitational field of mass $M$ that undergo Lense-Thirring precession due to the presence of the constant angular momentum of the source $J_0$.

\subsection{Perturbed spherical orbits}\label{ss2.3}

We now let $J=J_0 +J_1 t$ with $J_1\neq 0$ and consider the solution of equations \eref{eq:14}-\eref{eq:17} to linear order of perturbation beyond an arbitrary spherical orbit. Thus let
\begin{equation}\label{eq:28}
r=\rho (1+f),\quad \theta =\vartheta+g,\quad \phi=\varphi +h,
\end{equation}
where $f(t)$, $g(t)$ and $h(t)$ are all of $O(c^{-2})$ and depend linearly upon $J_1$. Substituting \eref{eq:28} in the equations of motion, we find
\begin{eqnarray}\label{eq:29}
\ddot{f}-2\dot{\vartheta}\dot{g}-2\hat{C}\dot{h} -3\omega^2_K f-2\ddot{\vartheta} g=\frac{2GJ_1\hat{C}t}{c^2\rho^3},\\
\ddot{g}-2\ddot{\vartheta}\cot (2\vartheta)g+2\dot{\vartheta}\dot{f}-2\hat{C}\cot \vartheta \dot{h}=-\frac{4GJ_1\hat{C}t}{c^2\rho^3} \cot \vartheta,\label{eq:30}\\
\dot{h}+\frac{2}{\hat{C}}\ddot{\vartheta}g+\frac{2\hat{C}}{\sin ^2\vartheta} f=\frac{2GJ_1 t}{c^2\rho^3} +\frac{\hat{C}R_0}{\sin^2\vartheta}.
\label{eq:31}\end{eqnarray}
Here, $\hat{C}=\omega \cos i$ and $R_0$ is a dimensionless integration constant. While equations \eref{eq:29} and \eref{eq:30} have been respectively obtained directly from \eref{eq:14} and \eref{eq:15}, equation \eref{eq:31} is obtained from \eref{eq:17}, since the right-hand side of this equation vanishes under our perturbation conditions. Substituting for $\dot{h}$ in \eref{eq:29} and \eref{eq:30} using \eref{eq:31} leads to
\begin{eqnarray}\label{eq:32}
\ddot{f}+\left( \frac{4\hat{C}^2}{\sin ^2\vartheta} -3\omega^2_K\right) f-2 (\dot{\vartheta}\dot{g}-\ddot{\vartheta}g)=\frac{6GJ_1\hat{C}t}{c^2\rho^3}+\frac{2R_0\hat{C}^2}{\sin^2\vartheta},\\
\ddot{g}+2\ddot{\vartheta} [2\cot \vartheta -\cot (2\vartheta)] g+4\ddot{\vartheta} f+2\dot{\vartheta}\dot{f} =2R_0\ddot{\vartheta}.\label{eq:33}
\end{eqnarray}

Inspection of equations~\eref{eq:32} and \eref{eq:33} reveals that they have the following solutions:
\begin{eqnarray}\label{eq:34}
f(t)=\frac{6J_1\cos i}{Mc^2} \omega_Kt+2R_0,\\
g(t)=-\left( \frac{6J_1 \cos i}{Mc^2} \omega_K t^2+3R_0t-\Theta_0\right) \dot{\vartheta},
\label{eq:35}\end{eqnarray}
where $\Theta_0$ is an integration constant. With these formulas for $f$ and $g$, \eref{eq:31} can be simply integrated and the result is
\begin{equation}\label{eq:36}
h(t)=\frac{GJ_1t^2}{c^2\rho^3} -\frac{\hat{C}}{\sin^2\vartheta} \left( \frac{6J_1 \cos i}{Mc^2} \omega_Kt^2 +3R_0t-\Theta_0\right) +\Phi_0,
\end{equation}
where $\Phi_0$ is another constant of integration.

It is possible to assume that $R_0=\Theta _0=\Phi_0=0$ without any loss in generality. To see this, let $J_1=0$; then, it is straightforward to demonstrate, using equations~\eref{eq:28} and \eref{eq:34}-\eref{eq:36}, that the resulting orbit $(r,\theta ,\phi)$ is simply a new \emph{spherical} orbit of constant radius $\rho (1+2R_0)$. It follows that one can set the dimensionless constants $R_0$, $\Theta_0$ and $\Phi_0$ equal to zero with no loss in generality. Then, the perturbation $(f,g,h)$ due to $J_1\neq 0$ is clearly secular leading to the instability of spherical orbits; in fact, these orbits spiral outward for $J_1\cos i>0$ and inward for $J_1\cos i<0$. These results provide independent confirmation of the some of the conclusions regarding perturbations of Keplerian orbits presented in \cite{4}, since a spherical orbit in the present treatment is simply a Lense-Thirring precessing circular orbit. A detailed investigation verifies this correspondence for the secular terms. We note, in this connection, that for $J_1=0$, the perturbed circular Keplerian orbit in \cite{4} differs from a spherical orbit by a harmonic term in the radial perturbation---see equation (71) of \cite{4}.

Finally, let us remark that the speed of the motion along the perturbed orbit is given by
\begin{equation}\label{eq:37}
v^2=v^2_0 - \frac{8GJ_1 \cos i}{c^2\rho} \omega_Kt,
\end{equation}
where $v_0$ is the speed of the unperturbed spherical orbit
\begin{equation}\label{eq:38}
v_0=\rho \omega \left( 1-\frac{J_0\omega_K}{Mc^2} \cos i\right).
\end{equation}
It follows from \eref{eq:37} that the motion is slower for $J_1\cos i>0$ and faster for $J_1 \cos i<0$. This confirms one of the main results of \cite{4} by a completely different analysis. On these physical grounds, we expect that our special solution contains the
\emph{dominant} secular terms of the general solution of the linear perturbation
equations.

\section{Extension of the Lense-Thirring approach}\label{s3}

We turn now to the gravitomagnetic perturbations of Keplerian ellipses, first treated in general by Lense and Thirring~\cite{1,2}. In this section, we extend their analysis to take into account the temporal variation of the gravitomagnetic field.

The method that we employ here is not a generalization of the approach
developed in the previous section for circular orbits. That is, rather than
directly perturbing a Lense-Thirring precessing Keplerian ellipse, we
follow, for the sake of simplicity, the linear perturbation method of Lense
and Thirring \cite{1,2} in the case of $J(t) = J_0 + J_1 t$.

It is possible to express the three second-order equations of motion \eref{eq:7} in terms of six first-order Lagrange planetary equations \cite{8}. If at any instant of time $t$ the perturbing force $\mathbf{F}$ is turned off, the test particle follows an osculating Keplerian ellipse about the central source. Thus instead of the position and velocity of the particle at time $t$, the state of the particle can be equally well characterized by the six orbital elements of the osculating ellipse at time $t$. The motion can therefore be described in terms of the evolution of the parameters of the instantaneous osculating ellipse. These parameters can be chosen in various ways; we employ the Delaunay action-angle variables $(\tilde{L},\tilde{G},\tilde{H}, \tilde{\ell}, \tilde{g}, \tilde{h})$ given by \cite{9}
\begin{eqnarray}\label{eq:39}
\tilde{L} = a^{1/2}, \quad \tilde{G}=[GMa(1-e^2)]^{1/2},\quad \tilde{H}=\tilde{G}\cos i,\\
\eqalign{\tilde{\ell} =u-e\sin u,\quad \tilde{g}=\mbox{argument of the pericenter},\cr
\quad \tilde{h}=\mbox{longitude of the ascending node}.\label{eq:40}}
\end{eqnarray}
Here, $a$ is the semimajor axis of the osculating ellipse, $e$ is its eccentricity, $i$ is the orbital inclination, $u$ is the eccentric anomaly and $\tilde{\ell}$ is the mean anomaly. The quantity $\tilde{G}$ is the magnitude of the specific orbital angular momentum vector $\tilde{\mathbf{G}}=\mathbf{x}\times \mathbf{v}$, while its $z$-component is denoted by $\tilde{H}$. Moreover, the radial position of the test particle along the osculating ellipse is given by
\begin{equation}\label{eq:41}
r=\frac{a(1-e^2)}{1+e\cos \tilde{v}},\quad r=a(1-e\cos u),
\end{equation}
where $\tilde{v}$ is the true anomaly.

The equations of motion are
\begin{eqnarray}\label{eq:42}
\fl \frac{d\tilde{L}}{dt} = \frac{\tilde{L}^3}{\tilde{G}} [F_r e \sin \tilde{v} +F_s (1+e\cos \tilde{v})],\\
\fl \frac{d\tilde{G}}{dt} = rF_s,\label{eq:43}\\
\fl \frac{d\tilde{H}}{dt} =r[F_s \cos i-F_n \sin i\cos (\tilde{v}+\tilde{g})],\label{eq:44}\\
\fl \frac{d\tilde{\ell}}{dt} = \omega_K +\frac{r}{\omega_Ka^2e} [F_r (-2e+\cos \tilde{v}+e\cos ^2\tilde{v}) -F_s (2+e\cos \tilde{v})\sin \tilde{v} ],\label{eq:45}\\
\fl \frac{d\tilde{g}}{dt} =-\frac{rF_n}{\tilde{G}} \cot i\sin (\tilde{v} +\tilde{g})+\frac{(1-e^2)^{1/2}}{\omega_Kae}\left( -F_r \cos \tilde{v} + F_s \frac{2+e\cos \tilde{v}}{1+e\cos \tilde{v}} \sin \tilde{v}\right),\label{eq:46}\\
\fl \frac{d\tilde{h}}{dt} = \frac{rF_n}{\tilde{G}} \frac{\sin (\tilde{v}+\tilde{g})}{\sin i} ,\label{eq:47}
\end{eqnarray}
where the Keplerian frequency is given by $\omega_K=(GM)^{1/2} /\tilde{L}^3$. Here, the perturbing force is given in terms of its radial, sideways and normal components,
\begin{equation}\label{eq:48}
\mathbf{F}=F_r\hat{\mathbf{r}}+F_s \hat{\mathbf{s}} +F_n\hat{\mathbf{n}}.
\end{equation}
That is, $\hat{\mathbf{r}}$ is the radial unit vector as before, $\hat{\mathbf{s}}=\hat{\mathbf{n}}\times \hat{\mathbf{r}}$ and $\hat{\mathbf{n}}$ is given by $\tilde{\mathbf{G}}=\tilde{G} \hat{\mathbf{n}}$; hence, $\hat{\mathbf{r}}$ and $\hat{\mathbf{s}}$ are in the plane of the osculating ellipse, while $\hat{\mathbf{n}}$ is normal to it.

To find the components of the perturbing force, it is useful to recall that the position vector of the test particle along the osculating ellipse has components
\begin{eqnarray}\label{eq:49}
x=r[\cos \tilde{h} \cos (\tilde{v} +\tilde{g} )-\sin \tilde{h} \cos i\sin (\tilde{v}+\tilde{g})],\\
y=r[\sin \tilde{h} \cos (\tilde{v} +\tilde{g}) +\cos \tilde{h} \cos i \sin (\tilde{v} +\tilde{g})],\label{eq:50}\\
z=r\sin i\sin (\tilde{v}+\tilde{g}),\label{eq:51}
\end{eqnarray}
while its velocity is given by
\begin{equation}\label{eq:52}
\mathbf{v} =\dot{r} \hat{\mathbf{r}}+\frac{\tilde{G}}{r} \hat{\mathbf{s}},\quad \dot{r} =GM \frac{e\sin \tilde{v}}{\tilde{G}}.
\end{equation}
Furthermore, with respect to the background $(x,y,z)$ coordinate system
\begin{eqnarray}\label{eq:53}
\hat{\mathbf{n}} =(\sin \tilde{h} \sin i,-\cos \tilde{h} \sin i,\cos i),\\
\eqalign{ \hat{\mathbf{s}} =(-\cos \tilde{h}\sin (\tilde{v}+\tilde{g} )-\sin \tilde{h} \cos i\cos (\tilde{v} +\tilde{g}),\cr
\hspace{.3in} -\sin \tilde{h} \sin (\tilde{v}+\tilde{g} )+\cos \tilde{h} \cos i\cos (\tilde{v}+\tilde{g}),\sin i\cos (\tilde{v}+\tilde{g})).\label{eq:54}}
\end{eqnarray}

The perturbing force \eref{eq:8} contains small relativistic terms and in our perturbation scheme the influence of these perturbing accelerations are simply additive. Therefore, we will ignore the post-Newtonian gravitoelectric terms in $\mathbf{F}$ and concentrate instead on the gravitomagnetic terms, namely,
\begin{equation}\label{eq:55}
\mathbf{F}'=\frac{2G}{c^2} \frac{\dot{\mathbf{J}} \times \mathbf{x}}{r^3} -\frac{2}{c} \mathbf{v}\times \mathbf{B}.
\end{equation}
We find that
 \begin{eqnarray}\label{eq:56}
\fl F'_r = \frac{2G(J_0+J_1 t) \tilde{H}}{c^2r^4},\\
\fl F'_s = \frac{2G\cos i}{c^2r^2} \left [ J_1 -(J_0 +J_1 t)\frac{GMe\sin \tilde{v}}{r\tilde{G}} \right],\label{eq:57}\\
\fl F'_n = -\frac{2G\sin i}{c^2r^2} \Big\{ J_1 \cos (\tilde{v} +\tilde{g})+(J_0 +J_1t)\frac{GM}{r\tilde{G}} [e\sin \tilde{g} - (2+3e\cos \tilde{v} )\sin (\tilde{v}+\tilde{g})]\Big\}.\label{eq:58}
\end{eqnarray}
With these perturbing functions, we can simply integrate \eref{eq:42}-\eref{eq:47} following the linear perturbation approach adopted by Lense and Thirring~\cite{1,2}. That is, we regard the orbital elements appearing on the right-hand side of the equations of motion as constants and employ $d\tilde{v}/dt=\tilde{G}/r^2$ for the osculating ellipse at time $t$; in fact, $\tilde{v}(t)$ and $t(\tilde{v})$ are obtained from
\begin{equation}\label{eq:59}
\frac{d\tilde{v}}{dt}=\frac{\omega_K}{(1-e^2)^{3/2}} (1+e\cos \tilde{v})^2.
\end{equation}
Assuming, for the sake of simplicity, that $t=0$ at $\tilde{v}=0$, we find
\begin{eqnarray} \label{eq:60}
\tilde{v}=\omega_Kt+2e\sin \omega_K t+\frac{5}{4} e^2 \sin 2\omega_Kt+O(e^3),\\
\omega_Kt=\tilde{v}-2e\sin\tilde{v}+\frac{3}{4} e^2 \sin 2\tilde{v}+O(e^3).
\label{eq:61}\end{eqnarray}
The integration of the equations of motion is now straightforward; we recover the Lense-Thirring results for $J_0$ and find new terms proportional to $J_1$. The latter terms are generally secular, of course, and render the Lense-Thirring precessions time-dependent. In connection with orbital instability, the rest of this section is devoted to the secular variation of $(a,e,i)$, which remained on average unchanged in the Lense-Thirring treatment.

It follows from \eref{eq:42} that
\begin{equation}\label{eq:62}
\frac{da}{dt}=\frac{4GJ_1 a(1-e^2)^{1/2} \cos i}{c^2\omega_K r^3} ,
\end{equation}
which can be integrated using \eref{eq:59} and the result is
\begin{equation}\label{eq:63}
\Delta a=\frac{4J_1 a\cos i}{Mc^2(1-e^2)} \Delta (\tilde{v}+e\sin \tilde{v}).
\end{equation}
Thus the semimajor axis and hence the Newtonian energy of the osculating ellipse will have a secular variation in time depending upon the sign of $J_1\cos i$, so that essentially all orbits are unstable for $J_1\neq 0$. Similarly, from $\tilde{G}^2=GMa(1-e^2)$, \eref{eq:42} and \eref{eq:43}, we find that the change in eccentricity is given by
\begin{equation}\label{eq:64}
\Delta e=\frac{2\omega_K\cos i}{Mc^2 (1-e^2)^{1/2}} [-J_0\Delta (\cos \tilde{v})+J_1 \Delta I] +\frac{2J_1\cos i}{Mc^2} \Delta I',
\end{equation}
where
\begin{equation}\label{eq:65}
I=\int^{\tilde{v}}_0 t(\lambda) \sin \lambda\, d\lambda ,\quad I'=\int^{\tilde{v}}_0 \frac{2\cos \lambda +e(1+\cos^2\lambda)}{1+e\cos \lambda}\, d\lambda .
\end{equation}
These integrals contain secular terms. For instance, relations \eref{eq:60} and \eref{eq:61} may be used to show that for $\tilde{v}=0$ at $t=0$,
\begin{equation}\label{eq:66}
\omega_K I=\sin \tilde{t} -\tilde{t}\cos \tilde{t} +\frac{e}{2} (\sin 2\tilde{t} -2\tilde{t} \cos 2\tilde{t})+O(e^2),
\end{equation}
where $\tilde{t}=\omega_Kt$. Integral $I'$ can be evaluated exactly (see, for instance, formula 2.553 3 on page 148 of \cite{10}); it can also be expressed in powers of $e$ as
\begin{equation}\label{eq:67}
I'=2\sin \tilde{v}+\frac{e}{4} (2\tilde{v}-\sin 2\tilde{v})+O(e^2).
\end{equation}
For an orbit with initial eccentricity much less than unity, the secular term that
is independent of the eccentricity in $\Delta e$ is given by the variation of
\begin{equation}\label{eq:68}
-2 \frac{J_1 \cos i}{Mc^2} \omega_Kt\cos \omega_Kt.
\end{equation}
Let us also note that $\Delta e$ and $\Delta a$ are both proportional to $\cos i$, so that an osculating polar orbit tends to preserve its shape.

%\begin{figure}
%\begin{center}{\includegraphics[width=4in,height=1.5in]{a.eps}}\\
%{\includegraphics[width=4in,height=1.5in]{e.eps}}\\
%{\includegraphics[width=4in,height=1.5in]{i.eps}}\\
%{\hspace{0.1in}\includegraphics[width=3.92in,height=1.6in]{v.eps}}
%\end{center}
%\caption{Plots of $a/a_0$, $e$, $i$ and $v$ of the osculating ellipse, respectively from the top panel down, versus $\omega_K^0t $ for the parameters given in the last paragraph of section 3.}
%\end{figure}
\begin{figure}
\centerline{\includegraphics[]{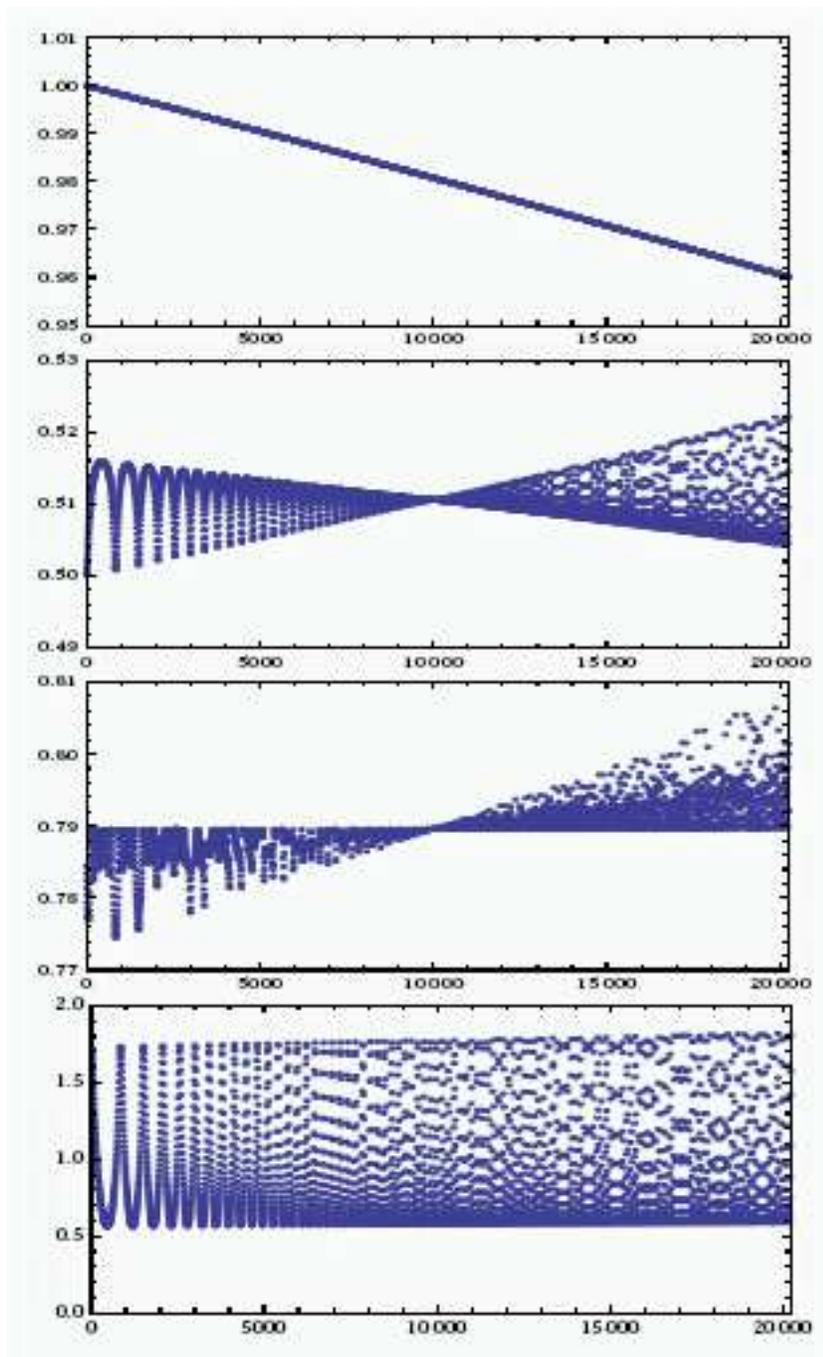}}
\caption{Plots of $a/a_0$, $e$, $i$ and $v$ of the osculating ellipse, respectively from the top panel down, versus $\omega_K^0t $ for the parameters given in the last paragraph of section 3.}
\end{figure}

Finally, equations \eref{eq:43} and \eref{eq:44} can be used to show that
\begin{equation}\label{eq:69}
\frac{di}{dt}=\frac{rF'_n}{\tilde{G}} \cos (\tilde{v}+\tilde{g}),
\end{equation}
where $F'_n$ is given by \eref{eq:58}. A simple integration reveals that 
\begin{equation}\label{eq:70}
\Delta i=-2 \frac{J_1 \sin i}{Mc^2} \Delta \mathcal{J} +2 \frac{\omega_K\sin i}{Mc^2(1-e^2)^{3/2}} \Delta \mathcal{J}',
\end{equation}
where
\begin{eqnarray}\label{eq:71}
\fl \mathcal{J} =\int^{\tilde{v}}_0 \frac{\cos ^2 (\lambda +\tilde{g})}{1+e\cos \lambda} d\lambda ,\\
\fl \mathcal{J} '=\int^{\tilde{v}}_0 (J_0+J_1 t) \Big[ (1+\frac{3}{2} e\cos \lambda )\sin (2\lambda +2\tilde{g} ) -e\sin \tilde{g}\cos (\lambda +\tilde{g}) \Big ] d\lambda .\label{eq:72}
\end{eqnarray}
Here, $\mathcal{J}$ can be evaluated exactly \cite{10}; moreover, $\mathcal{J}'$ can be expressed in powers of $e$ using \eref{eq:60} and \eref{eq:61}. We note that $\Delta i$ is proportional to $\sin i$, so that an equatorial orbit stays in the equatorial plane. In equations~\eref{eq:70} and \eref{eq:72}, the part proportional to $J_0$ contains only harmonic terms, while the part proportional to $J_1$ contains secular terms. For $e\ll 1$, the secular term that
is independent of the eccentricity in \eref{eq:70} is given by the variation of 
\begin{equation}\label{eq:73}
-2 \frac{J_1 \sin i}{Mc^2} \omega_Kt\cos^2 (\omega_Kt+\tilde{g}).
\end{equation}

To further illustrate orbital instability, we integrate numerically the equations of
motion for an initially eccentric Keplerian ellipse.  In practice, it turns
out to be simpler to integrate equation~\eref{eq:7}---where $\mathbf{F}$ is now replaced by $\mathbf{F}'$ given in \eref{eq:55}---with initial position
and velocity given by~\eref{eq:49}--\eref{eq:52}. Moreover, we employ the equations of motion
in dimensionless form, so that all lengths are given in units of $a_0$, the
initial semimajor axis, while time is given in units of $1/\omega^0_K$, the inverse of the initial Keplerian frequency. The equations of motion then depend on
two dimensionless parameters
\begin{equation}\label{eq:74}
\delta_0=\frac{2 J_0 \omega_K^0}{Mc^2},\qquad \delta_1=\frac{2 J_1}{Mc^2}.
\end{equation}

For the numerical results illustrated in figure 1, we choose $\delta_0=10^{-2}$. This corresponds approximately to an initial orbit of semimajor axis $a_0=40$ km
around a neutron star of mass $ M\approx 2 M_\odot$ and radius $\approx 10$ km      with a proper
rotation period of a millisecond. Furthermore, we choose $\delta_1=-10^{-6}$, so that after about 1600 Keplerian periods, the angular momentum of the
source decreases to zero. The integration is carried out for $\omega_K^0 t:0\to 20000$
such that $J(t):J_0\to -J_0$. This relatively rapid decrease of
angular momentum has been adopted here for the sake of illustration; in
fact, neutron stars generally lose angular momentum very slowly due to
electromagnetic braking torques.
     In figure 1, we plot  $a/a_0$, $e$, $i$    and the speed of the
motion  $v$   versus   $\omega_K^0t$. For the initial conditions at
$t=0$ and $\tilde v=0$, we choose $\tilde{h}_0 =\pi/6$, $\tilde{g}_0=\pi/3$ and $i_0=\pi/4$.
The initial eccentricity is chosen to be $e_0=0.5$; based on our numerical work, similar results are
expected for other initial eccentricities. The simple linear behavior of $a(t)$
depicted in the first panel of figure 1 can be obtained from \eref{eq:63}; that is, for $\omega_K^0 t\gg 1$,  equation \eref{eq:63} implies that
\begin{equation}\label{eq:75}
a\approx a_0\Big( 1+\frac{2\delta_1\cos i_0}{1-e_0^2}\omega_K^0 t \Big),
\end{equation}
in agreement with our numerical results. The second and third panels of figure 1 depict the oscillatory character of the eccentricity and the inclination angle, respectively, as the angular momentum of the source monotonically decreases from $J_0$ to $-J_0$. The midpoint of integration when $J = 0$ is a prominent feature of these graphs. In fact, the amplitudes of the quasi-periodic oscillations appear to be proportional to $J$. Moreover, the inclination angle $i$ tends to oscillate  \emph{toward} the angular momentum vector of the source $\mathbf{J}$.  As indicated
by our extensive numerical work, and is evident from the last three panels
of figure 1, we have to expect complexity in the details of the motion,
which is probably chaotic; therefore, only the overall trends are meaningful
here. The last panel confirms the expectation that for $J_1 \cos i < 0$, $v$ on the average has an increasing trend with time, in general agreement with equation \eref{eq:37} for the circular case. That is, for $J_1 \cos i > 0$ ($J_1 \cos i < 0$), the orbit generally tends to spiral outward (inward) accompanied by a corresponding decrease (increase) in its average speed.

\section{Discussion}\label{s4}

We have studied the instability of bound Keplerian orbits induced by a time-varying gravitomagnetic field in the post-Newtonian approximation. Circular and elliptical orbits have been treated separately in sections~\ref{s2} and \ref{s3}, respectively. The results are expected to be of interest in the study of variable collapsed astrophysical systems.

\ack C. Chicone was supported in part by the grant NSF/DMS-0604331.

\end{document}